# Title: Large Scale Manufacture of Phase Pure Two-Dimensional Metallic MoS$_2$ Nanosheets

**Authors:** Ziwei Jeffrey Yang[1], Zhuangnan Li[1]*, James Moloney[1], Leyi Loh[1], John Walmsley[1], Jiahang Li[1], Lixin Liu[1], Han Zang[1], Han Yan[1], Soumya Sarkar[1], Yan Wang[1], and Manish Chhowalla[1]*

**Affiliations:**

[1]Department of Materials Science & Metallurgy, University of Cambridge, Cambridge, UK

*Corresponding authors. Email: zl450@cam.ac.uk, mc209@cam.ac.uk

**Abstract**

Metallic monolayered [or two-dimensional (2D)] MoS$_2$ nanosheets show tremendous promise for energy storage and catalysis applications. However, state-of-the-art chemical exfoliation methods require > 48 hours to produce milligrams of metallic 2D MoS$_2$. Further, chemically exfoliated MoS$_2$ nanosheets are a mixture of metallic (1T or 1T' ~ 50% – 70%) and semiconducting (2H ~ 30% – 50%) phases. Here, we demonstrate large-scale and rapid (> 600 grams per hour) production of purely metallic phase 2D MoS$_2$ (and WS$_2$, MoSe$_2$) nanosheets using microwave irradiation. Atomic resolution imaging shows 1T or 1T' metallic phase in basal plane – consistent with ~100% metallic phase concentration measured by X-ray photoelectron spectroscopy. The high 1T phase concentration results in the highest exchange current density of 0.175 ± 0.03 mA cm$^{-2}$ and among the lowest Tafel slopes (39 – 43 mV dec$^{-1}$) measured to date for the hydrogen evolution reaction. In supercapacitors and lithium-sulfur pouch cell batteries, record-high volumetric capacitance of 753 ± 3.6 F cm$^{-3}$ and specific capacity of 1245 ± 16 mAh g$^{-1}$ (at exceptionally low electrolyte to sulfur ratio = 2 μL g$^{-1}$), respectively, are obtained. Our method provides a practical pathway for producing high quality purely metallic phase 2D materials for high performance energy devices.

**Introduction**

The metallic 1T phase of $MoS_2$ is interesting for applications in catalysis[1-4] and energy storage[5,6]. However, to translate the excellent performance from proof-of-concept devices with small scale (milligram, mg) samples into practical demonstrations, large-scale (kilogram, kg) and rapid production of high-quality materials is required. Bulk crystals of metallic transition metal dichalcogenides (TMDs) have been recently synthesized.[7,8] However, the exfoliation of metallic layered crystals into atomically thin nanosheets is challenging because of the strong interactions between layers due to delocalization of electrons.[9]

Chemical exfoliation of TMDs with n-butyllithium (n-BuLi) in hexane has been demonstrated for synthesis of metallic nanosheets.[10,11] n-BuLi is one of the most commonly used organolithium compounds and is produced in the highest annual quantities (2000-3000 metric tons).[12] In addition, compared to other organolithium compounds and isomers such as t-butyllithium (t-BuLi), it is more chemically stable and less pyrophoric.[13] The yield of single layer nanosheets is nearly 100% with this method. However, two challenges have hampered the use of this method for large scale production of metallic nanosheets. The first is that the as-exfoliated nanosheets are a mixture of metallic 1T and semiconducting 2H phases – with the 1T phase concentration being 50 – 70% for typical reaction conditions.[3,11,14,15] The second is that the reaction is slow – requiring 48 – 72 hours for homogeneous synthesis of tens of milligrams of nanosheets.[10,11,16-18] These issues have limited their implementation into practical energy devices where large quantities (kg to ton scale) are required.[19]

Efforts to accelerate the chemical exfoliation process leads to deterioration of nanosheet properties. For example, increasing the n-BuLi concentration leads to over-lithiation, resulting in decomposition of $MoS_2$ into Mo metal and lithium sulfide ($Li_2S$).[20,21] Increasing reaction temperature can enhance reaction kinetics. However, due to the strong reducing and deprotonating effects of n-BuLi, synthesis is typically performed in hexane for chemical stability. This restricts the reaction temperature to the boiling point of hexane (66 °C) – limiting the reaction and 1T phase conversion rates.

In this work, we use localized microwave heating for fast synthesis of high purity metallic 1T phase TMDs such as $MoS_2$, $WS_2$, and $MoSe_2$. Previous reports on microwave chemical exfoliation (MWCE) of $MoS_2$ yielded relatively low concentration of metallic 1T phase even at long irradiation times.[22] This is because the starting 2H phase $MoS_2$ is a wide band gap semiconductor and therefore does not efficiently absorb microwaves.[23] Here, we use carbon black nanoparticles (weight percent = 25%, particle size 100 - 200 nm, see Methods and Fig. 1a) dispersed on $MoS_2$ flakes as microwave susceptors to create local hot zones and increase the local reaction temperature above the boiling point of hexane (see Supplementary Fig. 1,2). Compared to other microwave absorbers such as silicon carbide (SiC), carbon black is less dense and disperse uniformly. This leads to rapid phase transformation and exfoliation of macro-scale $MoS_2$ powder as can be seen by the solutions shown in Fig. 1b. The solutions contain atomically thin 1T phase $MoS_2$ nanosheets as observed in Fig. 1c, which shows typical atomic force microscopy (AFM) image of flakes with ~ 1 nm thickness.[11]

We performed high-angle annular dark-field scanning transmission electron microscopy (HAADF-STEM) imaging to confirm the presence of metallic phase. Fig. 1d and 1e show HAADF-STEM images of MoS$_2$ nanosheets synthesized by conventional CE and MWCE for comparison. Atomic resolution Z-contrast observed in the HAADF mode was used to identify the different 1T, 2H, and distorted metallic 1T (1T') phases in MoS$_2$ nanosheets (Fig. 1f-k).[24,25] HAADF-STEM image of CE MoS$_2$ (Fig. 1d) shows mixed 1T' and 2H phases – consistent with previous reports.[24] In contrast, the MWCE samples (Fig. 1e) only show the metallic 1T/1T' phases. As the metallic 1T and 1T' phases can only be differentiated in STEM, we refer to both metallic phases as 1T in discussions below.

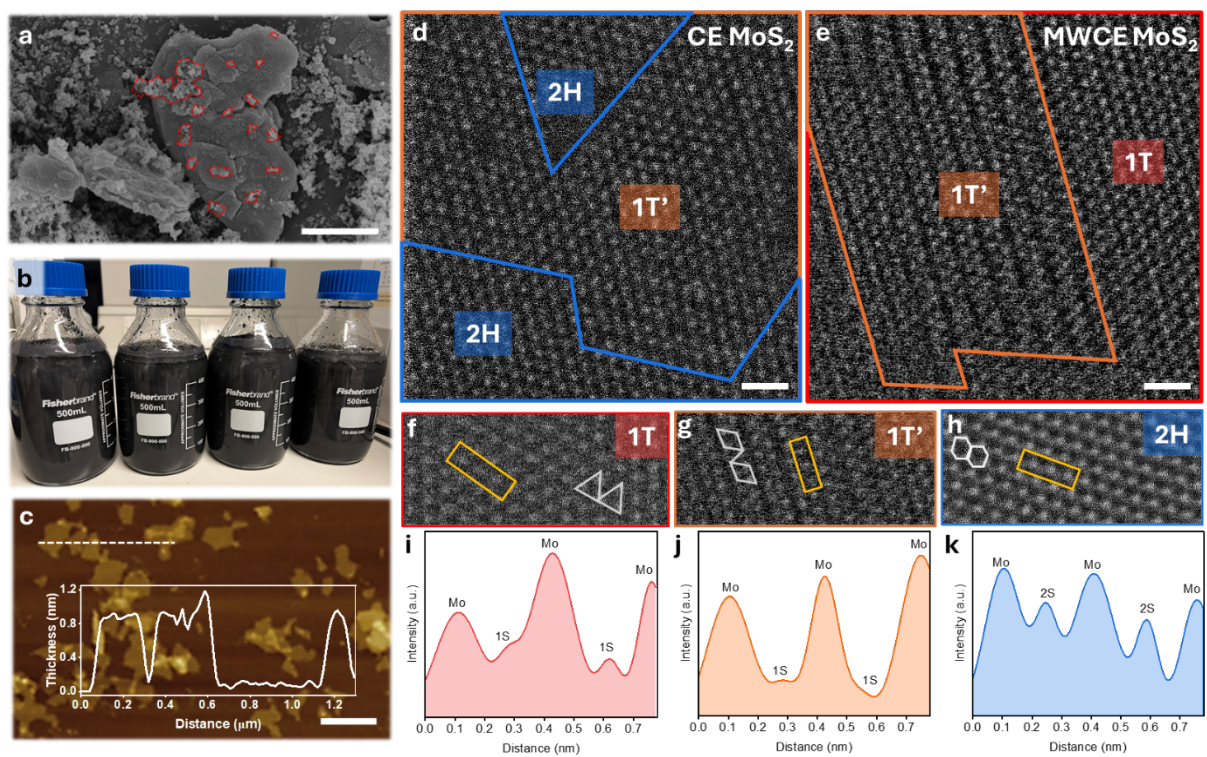

**Fig. 1: Imaging of CE and MWCE MoS$_2$.** **(a)** Scanning electron microscopy (SEM) image of precursor MoS$_2$ with ~ 100 nm carbon black susceptor nanoparticles (highlighted with red dotted lines). Scale bar = 1 μm. **(b)** Photograph of 100 g of MWCE MoS$_2$ nanosheets dispersed in deionized water. **(c)** AFM image of individual exfoliated MoS$_2$ nanosheets, and height profile along the dotted white line. Scale bar = 500 nm. **(d)** Typical HAADF-STEM image of CE MoS$_2$, showing mixed 2H and 1T' phases. **(e)** HAADF-STEM image of MWCE MoS$_2$, showing only 1T/1T' phases. **(f)** Enlarged 1T phase showing triangles that are visible due to high contrast Mo atoms. **(g)** Enlarged 1T' phase showing distorted 1T phase structure with diamond chain structure. **(h)** Enlarged 2H phase structure showing the hexagonal structure where both Mo and S atoms are visible. **(i, j, k)** Z-contrast intensity profiles along the atoms in yellow rectangles (in **f, g, h**) of 1T, 1T', and 2H phases. Sulfur signals in 1T/1T' phases are much lower relative to those in the 2H phase. Scale bars = 1 nm.

We then performed Raman mapping to assess the phase concentration uniformity in larger samples. Raman spectra of different types of MoS$_2$ are shown in Fig. 2a. Precursor 2H phase samples show the typical in-plane $E^1_{2g}$ and out-of-plane $A_{1g}$ modes of MoS$_2$ while the chemically exfoliated MoS$_2$ nanosheets show $J$ peaks that are characteristic of the 1T phase. In predominantly 1T phase samples, the $E^1_{2g}$ and $A_{1g}$ peaks are less pronounced than the $J$ peaks.[26] In addition, the CE sample shows the $J_2$ peak that originates from phase boundary between 1T and 2H phases.[27] In contrast, the $J_2$ peak is absent in MWCE samples and the $J_1$ and $J_3$ peaks are sharper and more prominent. Raman maps of 2H, CE, and MWCE MoS$_2$ nanosheets are shown in Fig. 2b-j. In the 2H phase precursor powder samples, only $E^1_{2g}$ and $A_{1g}$ peaks are observed (Fig. 2c, 2d). CE MoS$_2$ displays mostly 1T phase (Fig. 2f) with spots of 2H phase observed within the sample and at the edges of nanosheets where oxidation and reconstruction are more likely to occur (Fig. 2g). The MWCE samples show uniform 1T phase (Fig. 2i) with minute 2H phase concentration observed only at edges of the film and around a void (Fig. 2j).

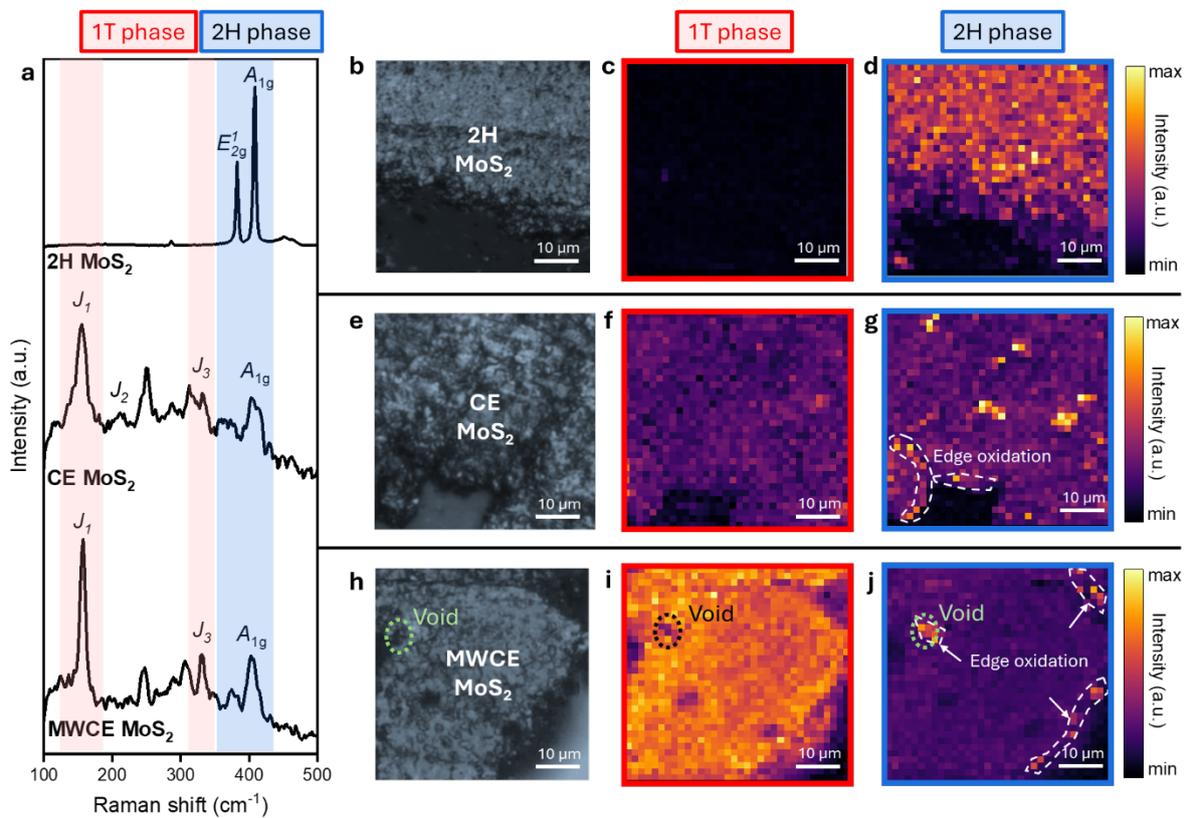

**Fig. 2: Raman spectroscopy of chemically exfoliated MoS$_2$ nanosheets.** (a) Typical Raman spectra of 2H, CE, and MWCE MoS$_2$ nanosheets. (b) Optical image, (c) $J_1$ and (d) $A_{1g}$ peak intensity maps of 2H MoS$_2$. (e) Optical image, (f) $J_1$ and (g) $A_{1g}$ peak maps of CE MoS$_2$. (h) Optical image, (i) $J_1$ and (j) $A_{1g}$ peak maps of MWCE MoS$_2$. The 2H phase-like features are only observed at edges in MWCE sample – in contrast to the CE sample.

X-ray photoelectron spectroscopy (XPS) was used to quantify the phase concentration and obtain chemical information about the MoS$_2$ nanosheets. XPS Mo $3d$ peaks are shown in Fig. 3a-c. The MoS$_2$ precursor powder shows signals of pure 2H phase with the usual native oxides. CE MoS$_2$ is a mixture of 1T and 2H phase. In contrast, MWCE MoS$_2$ consists of the 1T phase Mo $3d_{5/2}$ and $3d_{3/2}$ peaks (228.0 and 231.1 eV).[11] 2H phase fraction from native oxides and edges is ~26% in CE MoS$_2$, and ~ 4% in MWCE samples. Supplementary Fig. 3 displays Mo L$_3$ X-ray absorption spectra (XAS) for 2H, CE, MWCE MoS$_2$. Pre-edge shoulder at 2522.8 eV representing unoccupied Mo 4*d* states in the 1T phase is observed for both CE and MWCE MoS$_2$.

XPS line scans (Supplementary Fig. 4a) of CE and MWCE MoS$_2$ films over 5 mm are shown in Fig. 3d and 3e. For CE MoS$_2$, three different typical signals observed at various locations on the sample are labelled as ①, ②, and ③ in Fig. 3d. Location ① shows the 2H phase from unreacted precursors. The 227.8 eV peak at position ② is from elemental Mo formed due to over-lithiation of MoS$_2$. Location ③ shows the 1T phase peak. From the XPS analysis, the CE samples contain ~70% 1T phase and ~30% 2H phase. In contrast, the XPS line scan in Fig. 3e of MWCE samples shows very uniform 1T phase signal. Signals from positions ④, ⑤, and ⑥ are essentially the same and are representative of all the spectra measured along the line. Deconvolution of the Mo $3d_{5/2}$ peak for MWCE samples shows ~100% 1T phase with no identified decomposition to elemental Mo. Spectra from all points in the line scan are provided in Supplementary Fig. 4b and 4c.

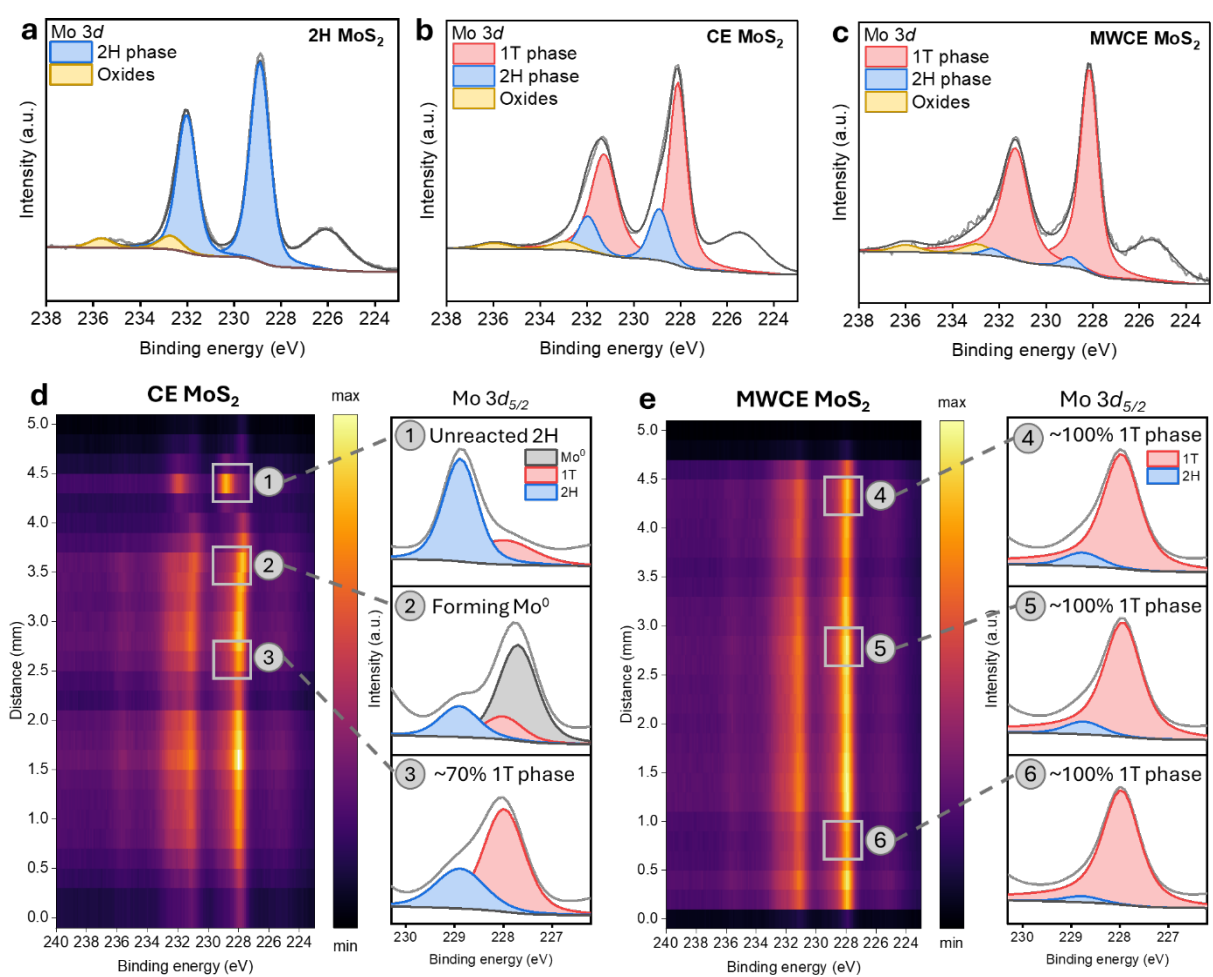

**Fig. 3: XPS characterization of MoS$_2$.** (**a**) Mo 3*d* XPS spectra of 2H (**b**) CE and (**c**) MWCE MoS$_2$. (**d**) Left: XPS line scan at equidistant points of CE MoS$_2$. Right: Typical spectra corresponding to unreacted 2H phase, elemental Mo, and mixture of 1T/2H phases along the line scan positions labelled as ①, ②, and ③, respectively. (**e**) Left: Similar line scan to (**d**) but for MWCE MoS$_2$. No unreacted 2H phase, elemental Mo, or mixture of 1T/2H phases are observed. Right: Typical spectra shown at locations ④,⑤, and ⑥ corresponding to pure 1T phase.

Our results show that MWCE enables large scale manufacture of 1T phase $MoS_2$. We examine the kinetics of 2H to 1T phase conversion by performing synthesis as a function of temperature (see analysis corresponding to Supplementary Fig. 5-7 for detailed description). With conventional CE, the 1T phase conversion rate reaches a maximum of 0.062 % $s^{-1}$ at the boiling point of hexane (66 °C), which is 51 times slower (reaction rate of ~ 3.2% $s^{-1}$) than for MWCE (> 96% 1T phase in 30 s). Based on this, the activation energy for the 2H → 1T phase transformation was determined to be 0.58 ± 0.06 eV from the Arrhenius plot of the rate constant ($k$) and temperature (Fig. 4a). The plot shows that the reaction in MWCE proceeds at an effective reaction temperature of 146 °C. That is, 80 °C above the boiling point of hexane. Despite this, we do not observe hexane evaporation because the global temperature of the mixture remains well below 66 °C. The short reaction time and large-scale production with MWCE translate to a production rate of 600 g $h^{-1}$, which is more than seven orders of magnitude faster than other synthesis methods (Fig. 4b). Manufacture of the same amount of material using conventional CE would take over 160 days. Recent report on gram-scale synthesis of metallic $MoS_2$ using plasma spray presents a fast synthesis route.[28] Our microwave is limited to 5 g batches, but we do not see any limitations to manufacture of larger batches with a larger reactor. MWCE also works for other transition metal dichalcogenides (TMDs). Raman and XPS spectra of MWCE synthesized 1T phase molybdenum selenide ($MoSe_2$) and tungsten disulfide ($WS_2$) are shown in Supplementary Fig. 8 and 9.

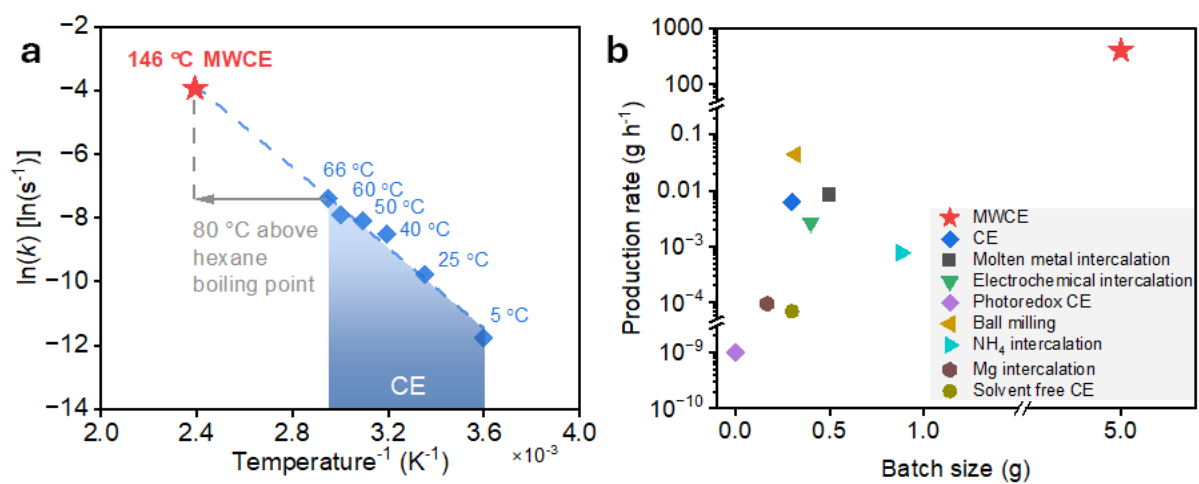

**Fig. 4: Reaction kinetics and scalability of MWCE. (a)** Arrhenius plot of phase conversion rate constant, $k$, at various temperatures for CE process. The slope (blue dotted line) yields an activation energy of 0.58 ± 0.06 eV per formula unit $MoS_2$ (56.1 ± 5.8 KJ $mol^{-1}$), the uncertainty originates from fitting of the slope. Extension of the line to the measured phase conversion rate for MWCE yields a local reaction temperature of 146 °C. **(b)** Comparison of MWCE production rate versus batch size for 1T phase $MoS_2$ with methods reported in literature (Supplementary Table 1).

Electrochemical properties of metallic $MoS_2$ are directly correlated with the 1T phase concentration. The MWCE samples show the highest electrical conductivity measured (Fig.

5a), which translates into excellent electrochemical performance. To this end, MWCE MoS$_2$ exhibits the highest exchange current densities for the hydrogen evolution reaction (HER) of any TMD catalyst, as shown in Fig. 5b. The exchange current density in HER indicates the intrinsic catalytic activity of a catalyst. Fig. 5c shows that MWCE MoS$_2$ catalysts exhibit a Tafel slope of 43 ± 4 mV dec$^{-1}$, which is close to the lowest values measured for MoS$_2$ catalysts (see Supplementary Table 2 for an exhaustive list of Tafel slopes from MoS$_2$ catalysts). The low Tafel slopes of MWCE MoS$_2$ catalysts suggest fast reaction kinetics due to the high electric conductivity and that Heyrovsky electrochemical desorption is the rate-determining step.

Supercapacitors with 70% 1T phase CE MoS$_2$ nanosheets show high volumetric capacitance of 650 F cm$^{-3}$ at a scan rate of 5 mV s$^{-1}$. [6] In comparison, supercapacitors based on MWCE MoS$_2$ show a record high volumetric capacitance of 753 ± 3.6 F cm$^{-3}$. The improved conductivity of MWCE 1T phase also improves the rate capability as shown by the near ideal rectangular cyclic voltammetry (CV) curves at various scan rates in Supplementary Fig. 10. MWCE MoS$_2$ shows a much better rate capability (60% capacitance retention at 500 mV s$^{-1}$) than that of the CE MoS$_2$ (35% capacitance retention at 500 mV s$^{-1}$) and various other 2D materials, as summarized in Fig. 5d.

Recently, we demonstrated that metallic MoS$_2$ is an excellent sulfur cathode host for Li-S battery cathodes. [5] In practical pouch cell batteries, MWCE MoS$_2$ cathodes with high areal sulfur loading (7.6 mg cm$^{−2}$) under exceptionally lean electrolyte conditions (electrolyte to sulfur ratio of 2.0 μL mg$^{−1}$) demonstrate a specific capacity of 1245 ± 16 mAh g$^{-1}$ (Supplementary Fig. 11). This is higher than that of CE MoS$_2$ cathodes under the same working conditions (1070 mAh g$^{-1}$), together with considerably higher cycling stability (Supplementary Fig. 11). Fig. 5f shows that compared to other Li-S cathodes in practical pouch cell batteries, MWCE MoS$_2$ demonstrates the highest specific capacity while maintaining the lowest electrolyte volume (Fig. 5e).

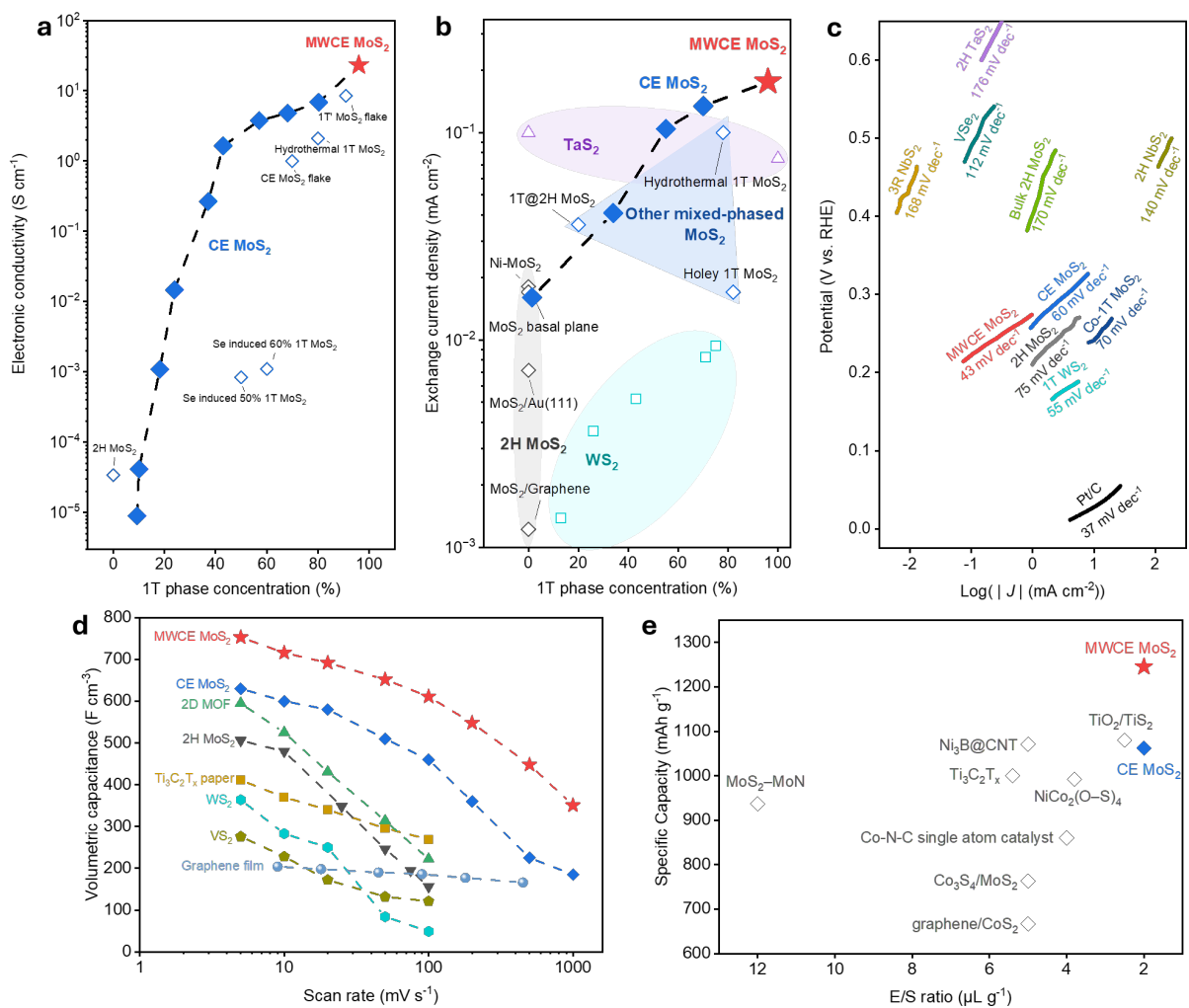

**Fig. 5: Electrochemical applications of MWCE MoS$_2$.** (a) Electronic conductivity as a function of 1T phase concentration for CE and MWCE MoS$_2$, and other reported values from literature. Data include mechanically exfoliated 2H MoS$_2$, 50% and 60% 1T MoS$_2$ induced by Se insertion[29], CE MoS$_2$ flake[11], 1T MoS$_2$ synthesized via hydrothermal method (hydrothermal 1T MoS$_2$), and 1T' MoS$_2$ flake[7]. Data for CE MoS$_2$ is taken from our previous study[3]. (b) Exchange current density as a function of 1T phase concentration for CE and MWCE MoS$_2$, and other TMD-based HER catalysts. Data include 2H MoS$_2$ edges on Au (111), graphene, vertically grown Ni-promoted 2H MoS$_2$ edges[30], 2H MoS$_2$ basal plane[31], 1T TaS$_2$[32], 2H TaS$_2$[33], WS$_2$ of various 1T phase concentrations[25], hydrothermal 1T MoS$_2$[34], holey 1T MoS$_2$[35], and 1T-incorporated 2H MoS$_2$ (1T@2H MoS$_2$)[36]. (c) Tafel slopes for CE and MWCE MoS$_2$, and other metallic TMD-based catalysts. Data include: 2H and 3R NbS$_2$[37], 2H TaS$_2$ and vanadium diselenide (VSe$_2$)[38], 1T WS$_2$[25], and Co-doped 1T MoS$_2$[39]. (d) Volumetric capacitance of MWCE MoS$_2$ and CE MoS$_2$ (ref. [6]), and other 2D-based supercapacitor electrodes (see Supplementary Table 3). (e) Specific capacity in relation to electrolyte to sulfur (E/S) ratio for practical Li-S pouch cell cathodes based on CE and MWCE MoS$_2$ and other values reported in literature (see Supplementary Table 4).

**Conclusion**

In summary, we have developed a microwave chemical exfoliation (MWCE) process for the rapid and scalable production of phase pure metallic $MoS_2$ nanosheets. Kinetic analyses of the phase transformation reveal that the extremely fast reaction is enabled by localized reaction temperature above the solvent boiling point. By using this method, 600 g h$^{-1}$ production rate for metallic $MoS_2$ nanosheets can be achieved. As-synthesized metallic $MoS_2$ nanosheets demonstrate excellent performance as electrodes for HER, supercapacitors, and Li-S batteries. Our research opens new opportunities for scalable synthesis of high-quality metallic 2D TMD materials.

## Methods

### Chemical exfoliation of MoS$_2$

Chemical exfoliation of MoS$_2$ was done similar to previous reports. Bulk 2H MoS$_2$ powder (0.3 g; Alfa Aesar) was immersed in hexane (15 mL; Sigma-Aldrich) and n-BuLi solution (1.6 M in hexane, 3 mL; Sigma-Aldrich). The mixture was refluxed for 48 hours under argon in an 80 °C oil bath. The Li$_x$MoS$_2$ product was then washed in hexane (3 × 50 mL) and exfoliated into nanosheets in deionized water or tetrahydrofuran (THF) at a concentration of 2mg mL$^{-1}$ via ultrasonication. Unreacted 2H precursor is removed via centrifuge after dispersion. Cumulated consumed energy is measured with an energy meter over the duration of the CE process. The daily energy usage for CE with a batch size of 0.3 g is 1.6 kWh.

### Microwave chemical exfoliation of MoS$_2$

Bulk 2H MoS$_2$ powder (0.3g) was mixed and grinded with porous carbon black powder (Nanografi EC-600JD, 0.1g) before immersing in hexane (15 mL) and n-BuLi solution (3 mL) in a liquid-tight PTFE vessel (MILESTONE SK-15 100 mL high pressure) under argon.

The vessel was irradiated under 2.450 GHz microwave for 30 s. The maximum microwave power was set to be 800W. A magnetic stirrer in the vessel was set to 240 RPM during irradiation. The temperature of the total mixture was controlled to not exceed 65 °C via an infrared temperature sensor (MILESTONE easyTEMP) from the bottom of the vessel. After cooling, the Li$_x$MoS$_2$ product is washed in hexane and exfoliated in deionized water or THF like that of chemically exfoliated MoS$_2$. Carbon black dispersion stability is poor in water and THF, can be removed from the MoS$_2$ nanosheets via high-speed centrifuge after dispersion and exfoliation. Cumulated consumed energy is calculated based on the power profile output by the instrument (Supplementary Fig. 12).

### Microwave chemical exfoliation of WS$_2$

Microwave exfoliation of WS$_2$ is similar to that of MoS$_2$. The bulk 2H MoS$_2$ powder is replaced with 2H WS$_2$ powder (0.3 g; Alfa Aesar).

### Microwave chemical exfoliation of MoSe$_2$

Microwave exfoliation of MoSe$_2$ is similar to that of MoS$_2$. The bulk 2H MoS$_2$ powder is replaced with 2H MoSe$_2$ powder (0.3 g; Alfa Aesar).

### Materials characterizations

STEM samples were prepared by drop-casting dilute MoS$_2$ nanosheet dispersions in deionized water onto lacey carbon grids. Aberration-corrected STEM measurements were conducted on a probe-corrected ThermoFisher Spectra 300 operating at 80 kV. Electron beam with a convergence angle of 24.7 mrad, beam current of 20 pA and dwell time of 2 μs were used during the acquisition. Collection semi-angles of 101-200 mrad are used to form the HAADF-STEM images. Crispen-Smooth filter[40] in the DigitalMicrograph software were applied to the STEM images to remove low-frequency amorphous hydrocarbon features and emphasize the high-frequency crystalline features for better clarity in structural analyses.

XPS (ThermoFisher Scientific NEXSA G2 using an Al Kα source with 50 μm probe size), SEM (FEI Nova NanoSEM, 5 kV beam accelerating voltage), and Raman spectroscopy (Renishaw InVia using a 514 nm laser beam). STEM imaging was performed using ThermoFisher Spectra 300 with an acceleration voltage of 80 kV. AFM was performed on a Bruker Icon system. XAS measurements were done at the I09 beamline at Diamond Light Source located at Dicot, UK. Exfoliated nanosheets were dispersed then dropped onto Cu foil in inert atmosphere before transferring into the analysis chamber. The incident angle was set to be -70 °.

**Electrochemical characterizations**

HER measurements were conducted using a three-electrode cell with 0.5 M argon-saturated sulfuric acid ($H_2SO_4$). CE and MWCE 1T $MoS_2$ was dispersed in deionized water and centrifuged at 18,000 RPM to remove the carbon susceptors, then freeze-dried and dispersed in deionized water. The 1T $MoS_2$ nanosheets were supported on glassy carbon working electrodes, similar to our previous reports.[25] 5 sets of working CE $MoS_2$ electrodes were annealed in vacuum at 100 °C for various time durations to achieve difference 1T phase concentrations. The working electrodes were allowed to cool in vacuum then tested immediately using an Ag/AgCl reference electrode and a graphite rod counter electrode. Linear sweep voltammetry was conducted with a scan rate of 5 mV s$^{-1}$. All potentials were normalized to reversible hydrogen electrode (RHE) using the equation

$$E_{RHE} = E_{Ag/AgCl} + E°_{AgCl} + 0.059V\, pH$$

where $E_{RHE}$ is the potential relative to RHE, $E_{Ag/AgCl}$ is the measured potential relative to the Ag/AgCl electrode, $E°_{Ag/AgCl}$ is taken to be 0.205 V at room temperature. Exchange current density is extracted from the Tafel slopes of various $MoS_2$ materials similar to our previous reports. [25]

Supercapacitor electrochemical measurements were conducted in a two-electrode electrochemical cell with 1.0 M argon-saturated sulfuric acid ($H_2SO_4$). MWCE 1T $MoS_2$ was dispersed in deionized water and centrifuged at 18,000 RPM to remove the carbon susceptors, then freeze-dried. 1 mg of freeze-dried MWCE 1T $MoS_2$ was mixed with 0.1 mg of PTFE in ethanol and loaded onto carbon paper with 15 mm diameter. Cyclic voltammetry data were collected in between 0 V and 0.80 V vs. counter electrode with various scan rates from 5 to 1000 mV s$^{-1}$.

**Li-S battery preparation**

CE or MWCE $MoS_2$ nanosheets and precipitated sulfur (Thermo Scientific) were mixed and annealed under Ar at 155 °C for 12 h at a 1:2.5 mass ratio to obtain $MoS_2$/S composite (71.4 wt % sulfur). The composite was then mixed with poly(vinylidene fluoride) (PVDF) (MTI Corporation) binder at a 95:5 mass ratio and made into a slurry with NMP. The slurry is coated on aluminum sheets. The areal sulfur loading is controlled to be 7.6 mg cm$^{-2}$. A thinner coating with a controlled areal loading of 1.75 mg cm$^{-2}$ is prepared for Li-S coin cells.

6 cm × 4.5 cm pouch cells are assembled in a dry room (relative humidity <0.1%). The $MoS_2$ cathode, a Celgard polypropylene separator, and a lithium foil anode (100 μm) are stacked and packed into Al-laminated films (MTI Corporation). Al and Ni tabs (MTI Corporation) were used for the outward connection of the cathode and anode, respectively. The N/P ratio of the cell is controlled to be 2. The pouch cell is then transferred into an Ar-filled glovebox for the injection of electrolyte and encapsulation. The E/S ratio is controlled to be 2.0 μL mg$^{-1}$. The pouch cells were cycled in the voltage range of 2.8–1.7 V at a rate of 0.1C for stability (1C = 1672 mAh g$^{-1}$). In coin cells, a Li foil is used as anode and the E/S ratio is controlled to be 30 μL mg$^{-1}$. The coin cells were cycled in the voltage range of 2.8–1.7 V at a rate of 0.1 - 1C for rate capability, and cyclic voltammetry data were collected in the same voltage window with various scan rates from 0.1 to 0.5 mV s$^{-1}$.

**Acknowledgements**

Funding: We acknowledge the support from the Faraday Institution LiSTAR program and characterization project (EP/S003053/1, FIRG014, and FIRG012). We also acknowledge Engineering and Physical Sciences Research Council (EPSRC H2CAT project, EP/V012932/1). We also acknowledge Royce Institute for access to the Royce Battery Suite, support of the Wolfson Electron Microscopy Suite and the use of the Thermo Fisher Spectra 300 (EP/R008779/1), and the Diamond light source for access to the I09 beamline (SI36790, SI39914). L.Loh and M.C. acknowledge the Royal Society Newton International Fellowship for the funding. Z.L. acknowledges the financial support and Research Fellowship from the Herchel Smith Fund and King's College, Cambridge.

**Author contributions:**

    Conceptualization: MC

    Methodology: ZJY, ZL, MC

    Investigation: ZJY, ZL, JM, L.Loh, JW, JL, L.Liu, HZ, HY, SS, YW

    Funding acquisition: MC

    Writing – original draft: MC

    Writing – review & editing: all authors

**Competing interests:** Authors declare that they have no competing interests.

**Data and materials availability:** All data are available in the main text or the supplementary materials.

**References**

1  Jaramillo, T. F. *et al.* Identification of active edge sites for electrochemical H2 evolution from MoS2 nanocatalysts. *Science* **317**, 100-102 (2007).
2  Li, Y. *et al.* MoS2 nanoparticles grown on graphene: an advanced catalyst for the hydrogen evolution reaction. *Journal of the American Chemical Society* **133**, 7296-7299 (2011).
3  Voiry, D. *et al.* Conducting MoS2 nanosheets as catalysts for hydrogen evolution reaction. *Nano Lett.* **13**, 6222-6227 (2013). https://doi.org/10.1021/nl403661s
4  Shi, Z. *et al.* Phase-dependent growth of Pt on MoS2 for highly efficient H2 evolution. *Nature* **621**, 300-305 (2023).
5  Li, Z. *et al.* Lithiated metallic molybdenum disulfide nanosheets for high-performance lithium–sulfur batteries. *Nat. Energy* **8**, 84-93 (2023). https://doi.org/10.1038/s41560-022-01175-7
6  Acerce, M., Voiry, D. & Chhowalla, M. Metallic 1T phase MoS2 nanosheets as supercapacitor electrode materials. *Nat. Nanotechnol.* **10**, 313-318 (2015). https://doi.org/10.1038/nnano.2015.40
7  Yu, Y. *et al.* High phase-purity 1T′-MoS2- and 1T′-MoSe2-layered crystals. *Nature Chemistry* **10**, 638-643 (2018). https://doi.org/10.1038/s41557-018-0035-6
8  Lai, Z. *et al.* Metastable 1T′-phase group VIB transition metal dichalcogenide crystals. *Nature Materials* **20**, 1113-1120 (2021). https://doi.org/10.1038/s41563-021-00971-y
9  Chhowalla, M. *et al.* The chemistry of two-dimensional layered transition metal dichalcogenide nanosheets. *Nat. Chem.* **5**, 263-275 (2013). https://doi.org/10.1038/nchem.1589
10  Dines, M. B. Lithium Intercalation via n-Butyllithium of the Layered Transition Metal Dichalcogenides. *Mat. Res. Bull.* **10**, 287-292 (1975).
11  Eda, G. *et al.* Photoluminescence from Chemically Exfoliated MoS2. *Nano Lett.* **11**, 5111-5116 (2011). https://doi.org/10.1021/nl201874w
12  Walfort, B. *et al.* [{(MeLi) 4 (dem) 1.5}∞] and [(thf) 3Li3Me {(NtBu) 3S}]—How to Reduce Aggregation of Parent Methyllithium. *Chemistry–A European Journal* **7**, 1417-1423 (2001).
13  Clayden, J. *Organolithiums: Selectivity for Synthesis*. (Pergamon, 2002).
14  Voiry, D., Mohite, A. & Chhowalla, M. Phase engineering of transition metal dichalcogenides. *Chem. Soc. Rev.* **44**, 2702-2712 (2015). https://doi.org/10.1039/c5cs00151j
15  Mei, L. *et al.* Metallic 1T/1T' phase TMD nanosheets with enhanced chemisorption sites for ultrahigh-efficiency lead removal. *Nat Commun* **15**, 7770 (2024). https://doi.org/10.1038/s41467-024-52078-y
16  Whittingham, M. S. & Dines, M. B. n–Butyllithium—An effective, general cathode screening agent. *Journal of the Electrochemical Society* **124**, 1387 (1977).
17  Lim, J. *et al.* Photoredox phase engineering of transition metal dichalcogenides. *Nature* **633**, 83-89 (2024). https://doi.org/10.1038/s41586-024-07872-5
18  Liu, L. *et al.* Phase-selective synthesis of 1T′ MoS2 monolayers and heterophase bilayers. *Nature Materials* **17**, 1108-1114 (2018). https://doi.org/10.1038/s41563-018-0187-1

19    Zhang, L. *et al.* Metal telluride nanosheets by scalable solid lithiation and exfoliation. *Nature* **628**, 313-319 (2024).
20    Mei, L. *et al.* Phase-switchable preparation of solution-processable WS2 mono- or bilayers. *Nature Synthesis*, 1-11 (2024).
21    Whittingham, M. S. CHEMISTRY OF INTERCALATION COMPOUNDS: METAL GUESTS IN CHALCOGENIDE HOSTS. *Prog. Solid St. Chem.* **12**, 41-99 (1978).
22    Benavente, E. & Gontilez, G. Microwave Activated Lithium Intercalation in Transition Metal Sulfides. *Mat. Res. Bull.* **32**, 709-717 (1997).
23    Wang, Z. *et al.* Carbon-enabled microwave chemistry: From interaction mechanisms to nanomaterial manufacturing. *Nano Energy* **85** (2021). https://doi.org/10.1016/j.nanoen.2021.106027
24    Eda, G. *et al.* Coherent Atomic and Electronic Heterostructures of Single-Layer MoS2. *ACS Nano* **6**, 7311-7317 (2012).
25    Voiry, D. *et al.* Enhanced catalytic activity in strained chemically exfoliated WS2 nanosheets for hydrogen evolution. *Nat. Mater.* **12**, 850-855 (2013). https://doi.org/10.1038/nmat3700
26    Jiménez Sandoval, S., Yang, D., Frindt, R. F. & Irwin, J. C. Raman study and lattice dynamics of single molecular layers of MoS2. *Physical Review B* **44**, 3955-3962 (1991). https://doi.org/10.1103/PhysRevB.44.3955
27    Tan, S. J. R. *et al.* Chemical Stabilization of 1T′ Phase Transition Metal Dichalcogenides with Giant Optical Kerr Nonlinearity. *Journal of the American Chemical Society* **139**, 2504-2511 (2017). https://doi.org/10.1021/jacs.6b13238
28    Kiran, P. S. *et al.* Scaling up simultaneous exfoliation and 2H to 1T phase transformation of MoS2. *Advanced Functional Materials* **34**, 2316266 (2024).
29    Jiang, L. *et al.* Se and O co-insertion induce the transition of MoS2 from 2H to 1T phase for designing high-active electrocatalyst of hydrogen evolution reaction. *Chemical Engineering Journal* **425**, 130611 (2021).
30    Tsai, C., Abild-Pedersen, F. & Nørskov, J. K. Tuning the MoS2 edge-site activity for hydrogen evolution via support interactions. *Nano letters* **14**, 1381-1387 (2014).
31    Voiry, D. *et al.* The role of electronic coupling between substrate and 2D MoS2 nanosheets in electrocatalytic production of hydrogen. *Nat. Mater.* **15**, 1003-1009 (2016). https://doi.org/10.1038/nmat4660
32    Huan, Y. *et al.* Vertical 1T–TaS2 synthesis on Nanoporous gold for high-performance electrocatalytic applications. *Advanced Materials* **30**, 1705916 (2018).
33    Shi, J. *et al.* Two-dimensional metallic tantalum disulfide as a hydrogen evolution catalyst. *Nature communications* **8**, 958 (2017).
34    Geng, X. *et al.* Pure and stable metallic phase molybdenum disulfide nanosheets for hydrogen evolution reaction. *Nature communications* **7**, 10672 (2016).
35    Yin, Y. *et al.* Contributions of phase, sulfur vacancies, and edges to the hydrogen evolution reaction catalytic activity of porous molybdenum disulfide nanosheets. *Journal of the American Chemical Society* **138**, 7965-7972 (2016).
36    Shi, S., Gao, D., Xia, B., Liu, P. & Xue, D. Enhanced hydrogen evolution catalysis in MoS 2 nanosheets by incorporation of a metal phase. *Journal of Materials Chemistry A* **3**, 24414-24421 (2015).

37	Yang, J. *et al*. Ultrahigh-current-density niobium disulfide catalysts for hydrogen evolution. *Nature Materials* **18**, 1309-1314 (2019). https://doi.org/10.1038/s41563-019-0463-8
38	Chia, X., Ambrosi, A., Lazar, P., Sofer, Z. & Pumera, M. Electrocatalysis of layered Group 5 metallic transition metal dichalcogenides (MX 2, M= V, Nb, and Ta; X= S, Se, and Te). *Journal of Materials Chemistry A* **4**, 14241-14253 (2016).
39	Jian, J. *et al*. Cobalt and Aluminum Co-Optimized 1T Phase MoS2 with Rich Edges for Robust Hydrogen Evolution Activity. *ACS Sustainable Chemistry & Engineering* **10**, 10203-10210 (2022). https://doi.org/10.1021/acssuschemeng.2c01836
40	Nixon, M. & Aguado, A. *Feature Extraction & Image Processing*. 2 edn, (Academic Press, 2008).